\documentclass[bm,aps,showpacs,amsmath,preprint,amssymb,12pt]{revtex4}
\usepackage{graphicx}
\usepackage{bm}

\begin{document}
{~}
\vspace*{1cm}

\title{Squashed Kerr-G\"odel Black Holes\\
  - {\it  Kaluza-Klein Black Holes with \\
  Rotations of Black Hole and Universe} - 
\vspace{1cm}
}
\author{Shinya Tomizawa, Hideki Ishihara, Ken Matsuno and Toshiharu Nakagawa}
\affiliation{ 
Department of Mathematics and Physics,
Graduate School of Science, Osaka City University,
3-3-138 Sugimoto, Sumiyoshi, Osaka 558-8585, Japan
\vspace{2cm}
}

\begin{abstract}
Applying {\it squashing transformation} to Kerr-G\"odel black hole solutions, we present a new type of a rotating Kaluza-Klein black hole solution to the five-dimensional Einstein-Maxwell theory with a Chern-Simon term. The new solutions generated via the squashing transformation have no closed timelike curve everywhere outside the black hole horizons. At the infinity, the metric asymptotically approaches a twisted $\rm S^1$ bundle over a four-dimensional Minkowski space-time. One of the remarkable features is that the solution has two independent rotation parameters along an extra dimension associated with the black hole's rotation and the G\"odel's rotation. The space-time also admits the existence of two disconnected  ergoregions, an inner ergoregion and an outer ergoregion. These two ergoregions can rotate in the opposite direction as well as in the same direction.
\end{abstract}

\preprint{}
\preprint{OCU-PHYS 294 \ AP-GR 56}
\pacs{04.50.+h  04.70.Bw}
\date{\today}
\maketitle

\section{Introduction}

The study of the five-dimensional Einstein-Maxwell theory with a Chern-Simon term plays a important role in elucidating the structure of the string theory since it is the bosonic sector of the minimal supergravity. All bosonic supersymmetric solutions of minimal supergravity in five dimensions are classified 
by Gauntlett {\it et al}~\cite{Gauntlett0}.
Many interesting supersymmetric (BPS) solutions to the five-dimensional Einstein-Maxwell equations (with a Chern-Simon term) have been found by a lot of authors.  
Following on this classification of the five-dimensional supersymmetric solutions, they have been constructed on hyper-K\"ahler 
base spaces.
The first asymptotically flat supersymmetric black hole solution, Breckenridge-Myers-Peet-Vafa (BMPV) solution, was constructed on the four-dimensional Euclid space~\cite{BMPV}. 
A supersymmetric black hole solution with a compactified extra dimension 
on the Euclidean self-dual Taub-NUT base space
was constructed by Gaiotto {\it et al}~\cite{Gaiotto}. 
It was extended to multi-black hole solution with the same asymptotic structure~\cite{IKMT}.
One of the most interesting properties is that the possible spatial topology of the horizon of each black hole is the lens space $L(n;1)={\rm S}^3/{\mathbb Z_{n}}$ ($n $:the natural numbers ) in addition 
to $\rm S^3$. 
Similarly, black hole solution on the Eguchi-Hanson space~\cite{IKMT2} was also constructed. Some supersymmetric black ring solutions have been also found, based on the 
construction of the solutions by Gauntlett {\it et al}~\cite{Gauntlett0}.  
Elvang {\it et al} found the first supersymmetric black ring solution 
with asymptotic 
flatness on the four-dimensional Euclidean base space, which is specified by three 
parameters, mass and two independent angular momentum components~\cite{Elvang}. 
Gauntlett and Gutowski also constructed a multi-black ring solution on the same 
base space~\cite{Gauntlett,Gauntlett2}. 
The BPS black rings with three arbitrary charges and three dipole charges on the flat space were also constructed in ~\cite{Elvang2,Bena3}. 
The BPS black ring solutions on the Taub-NUT base 
space~\cite{Bena,Bena2,Gaiotto2,Elvang2} and the Eguchi-Hanson space~\cite{TIKM,T}were constructed.

In recent years, some non-BPS black hole solutions have also been found in addition to supersymmetric black hole solutions.
Although no one has found higher-dimensional Kerr-Newman solutions in Einstein-Maxwell theory yet, Cvetic {\it et al}~\cite{CLP} found a non-extremal, charged  and rotating black hole solution with asymptotic flatness in the five-dimensional Einstein-Maxwell theory with a Chern-Simon term.
In the neutral case, the solution reduces to the same angular momenta case of the Myers-Perry black hole solution~\cite{Myers}. Exact solutions of non-BPS Kaluza-Klein black hole solutions are found in 
neutral case~\cite{DM,GW} and charged case~\cite{IM}. 
These solutions have a non-trivial asymptotic structure, i.e., 
they asymptotically approach a twisted $\rm S^1$ bundle over the four-dimensional 
Minkowski space-time. The horizons are deformed due to this non-trivial asymptotic 
structure and have a shape of a squashed $\rm S^3$, where $\rm S^3$ is regarded as 
a twisted bundle over a $\rm S^2$ base space. The ratio of the radius $\rm S^2$ to 
that of $\rm S^1$ is always larger than one.

As was proposed by Wang, a kind of Kaluza-Klein black hole solutions 
can be generated by the \lq {\it squashing transformation}\rq~ from 
black holes with asymptotic flatness~\cite{Wang}. 
In fact, he regenerated the five-dimensional Kaluza-Klein black hole solution found by Dobiasch and Maison~\cite{DM,GW} from the five-dimensional Myers-Perry black hole solution with two equal angular momentum\footnote{
The solution generated by Wang coincides with the solution in Ref.\cite{DM,GW}. }. In the previous work~\cite{NIMT}, applying the squashing transformation to the Cvetic {\it et al}'s charged rotating black hole solution~\cite{CLP} 
in vanishing cosmological constant case, 
we obtain the new Kaluza-Klein black hole solution in the five-dimensional 
Einstein-Maxwell theory with a Chern-Simon term. This is the generalization of the Kaluza-Klein black hole solutions in Ref.~\cite{DM,GW,IM}. This solution has four parameters, the mass, the angular momentum in the direction of an extra dimension, the electric charge and the size of the extra dimension. The solution describes the physical situation such that in general a non-BPS black hole is boosted in the direction of the extra dimension. As the interesting feature of the solution, unlike the static solution~\cite{IM}, the horizon admits a prolate shape in addition to a round $\rm S^3$.  The solution has the limits to the supersymmetric black hole solution and a new extreme non-BPS black hole solutions and a new rotating black hole solution with a constant twisted $\rm S^1$ fiber.

In this article, applying this squashing transformation to Kerr-G\"odel black hole solutions~\cite{Gimon-Hashimoto}, we constructed a new type of rotating Kaluza-Klein black hole solutions to the five-dimensional Einstein-Maxwell theory with a Chern-Simon term. We also investigate the features of the solutions. Though the G\"odel black hole solutions have closed timelike curves in the region away from the black hole, the new Kaluza-Klein black hole solutions generated by the squashing transformation have no closed timelike curve everywhere outside the black hole horizons. At the infinity, the space-time approaches a twisted $\rm S^1$ bundle over a four-dimensional Minkowski space-time. The solution has four independent parameters, the mass parameter, the size of an extra dimension and two kinds of rotations parameters in the same direction of the extra dimension. These two independent parameters are associated with the rotations of the black hole and the universe. In the case of the absence of a black hole, the solution describes the Gross-Perry-Sorkin (GPS) monopole which is boosted in the direction of an extra dimension and has an ergoregion by the effect of the rotation of the universe.

The rest of this article is organized as follows. 
In Sec.\ref{sec:solution}, we present a new Kaluza-Klein black hole solution in the five-dimensional Einstein-Maxwell theory with a Chern-Simon term. 
In Sec.\ref{sec:special}, we study the special cases of our solution. In Sec.\ref{sec:feature}, we investigate the basic features of the solution. In Sec.\ref{sec:summary}, we summarize the results in this article. In the Appendix, we will present more general solution.

\section{Squashed Kerr-G\"odel black hole}\label{sec:solution}
\subsection{Solution}
First, we present the metric of a new Kaluza-Klein black hole solution to the five-dimensional Einstein-Maxwell theory whose action is given by the action:
\begin{eqnarray}
S=\frac{1}{16\pi G_5}\int\left(R*1 - 2F\wedge *F - \frac{8}{3 \sqrt 3} F \wedge F \wedge A \right).
\end{eqnarray}
The metric and the gauge potential are given by
\begin{eqnarray}
ds^2=-f(r)dt^2-2g(r)\sigma_3dt+h(r)\sigma_3^2+\frac{k(r)^2dr^2}{V(r)}+\frac{r^2}{4}[k(r)(\sigma_1^2+\sigma_2^2)+\sigma_3^2],
\end{eqnarray}
and
\begin{eqnarray}
{\bm A}=\frac{\sqrt{3}}{2}jr^2\sigma_3,
\end{eqnarray}
respectively, where the functions in the metric are 
\begin{eqnarray}
&&f(r)=1-\frac{2m}{r^2},\\
&&g(r)=jr^2+\frac{ma}{r^2},\\
&&h(r)=-j^2r^2(r^2+2m)+\frac{ma^2}{2r^2},\\
&&V(r)=1-\frac{2m}{r^2}+\frac{8jm(a+2jm)}{r^2}+\frac{2ma^2}{r^4},\\
&&k(r)=\frac{V(r_\infty)r_\infty^4}{(r^2-r_\infty^2)^2}
\end{eqnarray}
and the left-invariant $1$-forms on $\rm S^3$ are given by
\begin{eqnarray}
&&\sigma_1=\cos\psi d\theta+\sin\psi\sin\theta d\phi,\\
&&\sigma_2=-\sin\psi d\theta+\cos\psi\sin\theta d\phi,\\
&&\sigma_3=d\psi+\cos\theta d\phi.
\end{eqnarray} 
The coordinates $r,\theta,\phi$ and $\psi$ run the ranges of $0<r<r_\infty$, $0\le \theta<\pi$, $0\le \phi<2\pi$, $0\le \psi<4\pi$, respectively. $m,a,j$ and $r_\infty$ are constants. The space-time has the timelike Killing vector fields $\partial_t$ and two spatial Killing vector fields with closed orbits, $\partial_\phi$ and $\partial_\psi$. In the limit of $k(r)\to 1$, i.e., $r_\infty\to\infty$ with the other parameters kept finite, the metric coincides with that of the Kerr-G\"odel black hole solution~\cite{Gimon-Hashimoto} with CTCs. 
In this article, we assume that the parameters $j,m,a$ and $r_\infty$ appearing in the solutions satisfy the following inequalities
\begin{eqnarray}
&&m>0,\label{eq:para1}\\
&&\frac{r^2_\infty}{m}>1-4j(a+2jm)>\sqrt{\frac{2}{m}}|a|,\label{eq:para2}\\
&&r_\infty^4-2m(1-4j(a+2jm))r^2_\infty+2ma^2>0,\label{eq:para3}\\
&&-4j^2r^6_\infty+(1-8j^2m)r_\infty^4+2ma^2>0.\label{eq:para4}
\end{eqnarray}
As will be explained later, these are the necessary and sufficient conditions that there are two horizons and no CTCs outside the horizons. Eqs. (\ref{eq:para1})-(\ref{eq:para3}) are conditions for the presence of two horizons, and Eq.(\ref{eq:para4}) is the condition for the absence of CTCs outside the horizons. It is noted that in the limit of $r_\infty\to\infty$ with the other parameters finite, Eq.(\ref{eq:para4}) can not be satisfied. Thus, applying the squashing transformation to the Kerr-G\"odel black hole solution, we can obtain such a Kaluza-Klein black hole solution without CTCs everywhere outside the black hole.

\subsection{Parameter region}

Here we derive the parameter region (\ref{eq:para1})-(\ref{eq:para4}) from the demands for the presence of two horizons and the absence of CTC outside the horizons.
As will be shown later, the Killing horizons are located at 
\begin{eqnarray}
r_\pm^2=\pm(1-4j(a+2jm))\pm \sqrt{m^2(1-4j(a+2jm))^2-2ma^2},
\end{eqnarray}
which are the roots of the equation $V(r)=0$, i.e.,  the quadratic equations with respect to $r^2$,
\begin{eqnarray}
v(r^2):=r^4-2m(1-4j(a+2jm))r^2+2ma^2=0.
\end{eqnarray}
This equation has two different real roots within the range of $(0,r_\infty^2)$ if and only if the parameters in the metric obey the inequalities
\begin{eqnarray}
&&v(0)=2ma^2>0,\\
&&v(r_\infty^2)=r_\infty^4-2m(1-4j(a+2jm))r^2_\infty+2ma^2>0,\\
&&r^2_\infty>m[1-4j(a+2jm)]>0,\\
&&m^2\left[1-4j(a+2jm)\right]^2-2a^2>0.
\end{eqnarray}
Under the condition $a\not= 0$, these inequalities are equivalent to (\ref{eq:para1})-(\ref{eq:para3}).

Next, in order to avoid the existence of CTCs outside the horizons, the parameters are chosen so that the two-dimensional $(\psi,\phi)$-part of the metric is positive-definite everywhere outside the outer horizon. To do so, it is sufficient to require $g_{\psi\psi}>0$ since $g_{\phi\phi}=g_{\psi\psi}\cos^2\theta+r^2k(r)\sin^2\theta/4>0$ is automatically assured if it is satisfied. Hence, the following inequality must be satisfied everywhere in the region $[r_+^2,r_\infty^2]$.
\begin{eqnarray}
g_{\psi\psi}=h(r)+\frac{r^2}{4}>0\Longleftrightarrow u(r^2):=-4j^2r^6+(1-8j^2m)r^4+2ma^2>0.
\end{eqnarray}
Here, it should be noted that it is sufficient to impose $u(r_\infty^2)>0$, which gives the equality (\ref{eq:para4}). Consequently, the condition for the absence of CTCs outside the horizons is equal to the demand for the absence of CTCs at the infinity.

\begin{figure}[!h]
 \vspace{1cm}
  \begin{center}
   \includegraphics[width=0.8\linewidth]{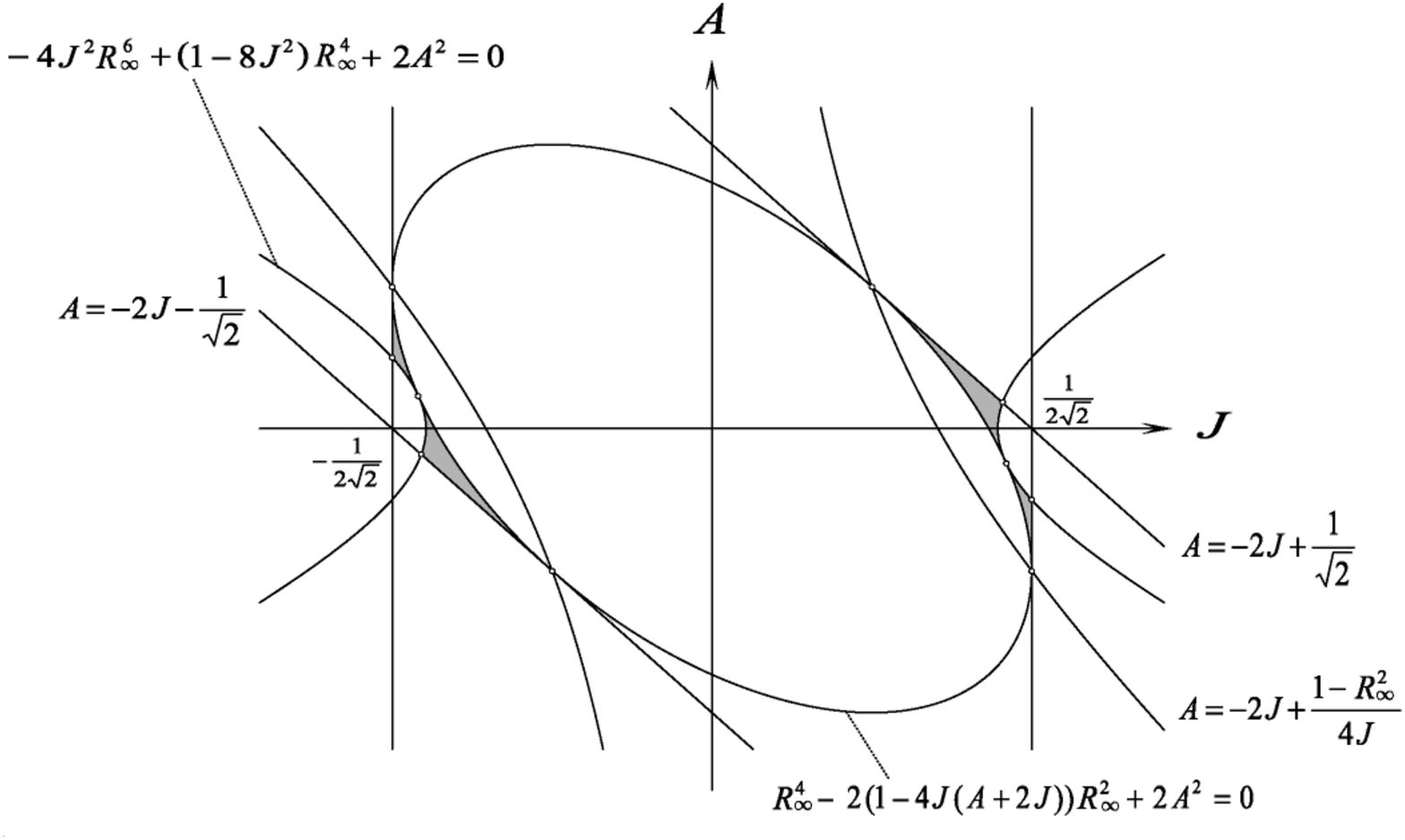}
 \caption{The parameter region in the $(J,A)$-plane in the case of 
$0<R^2_\infty<1$.  \label{fig:p1}}
  \end{center}
  \vspace{1cm}
  \begin{center}
   \includegraphics[width=0.8\linewidth]{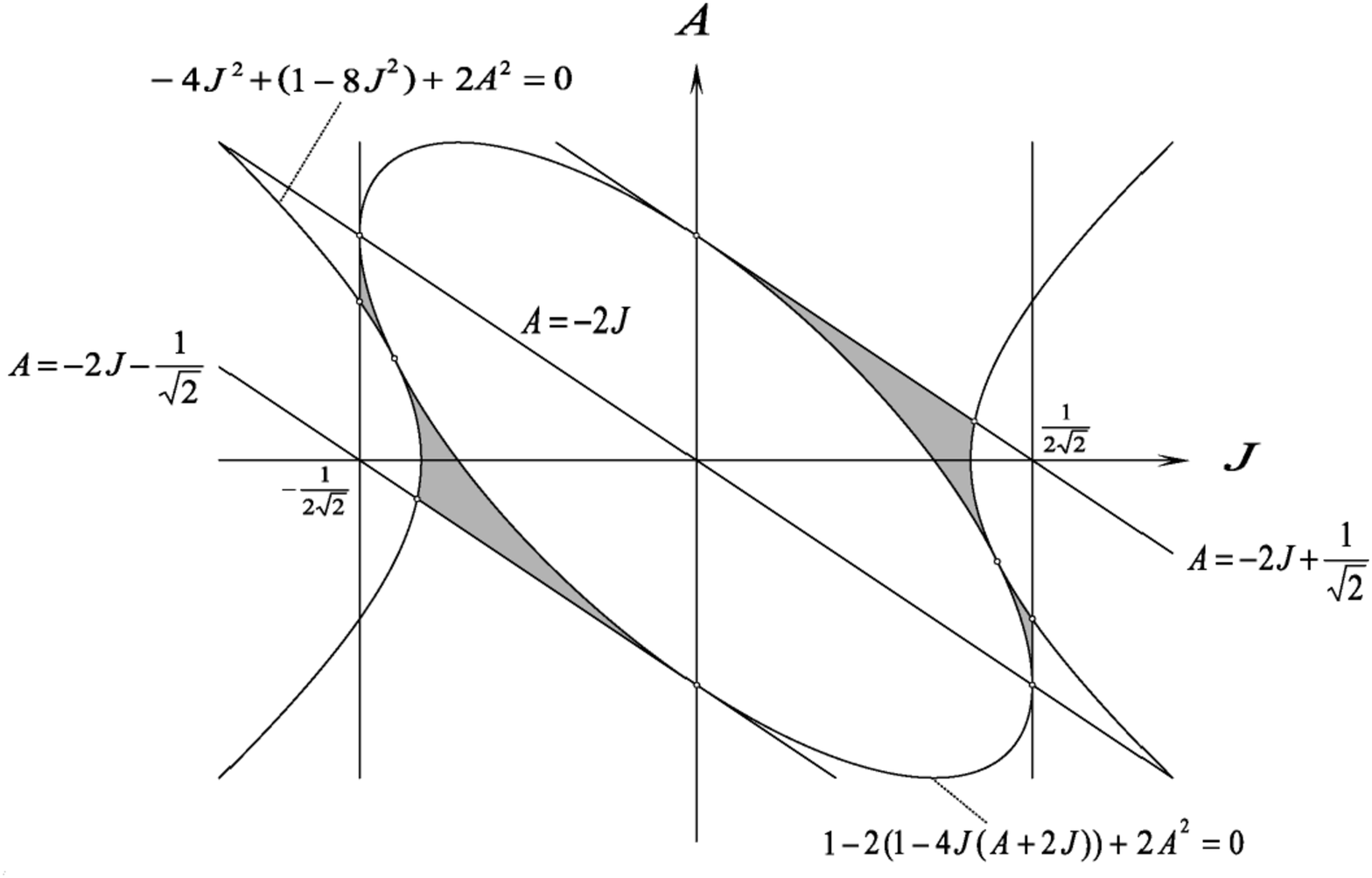}
  \caption{The parameter region in the $(J,A)$-plane in the case of $R^2_\infty=1$.  \label{fig:p2}}
  \end{center}
 \end{figure}

\begin{figure}[!h]
\begin{center}
\includegraphics[width=0.8\linewidth]{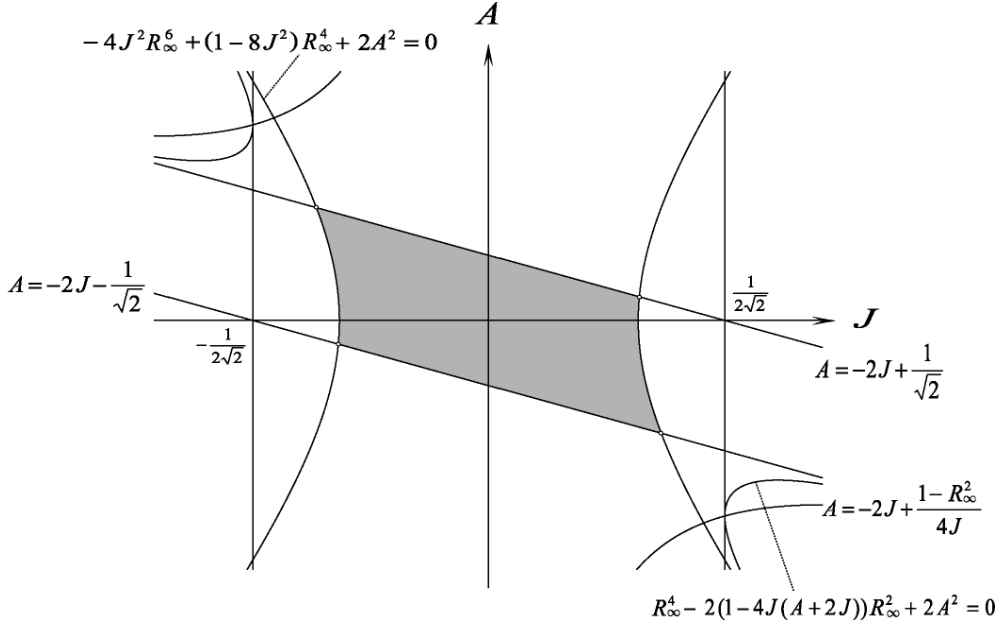}
\caption{The parameter region in the $(J,A)$-plane in the case of 
$2<R^2_\infty$.\label{fig:p5}
}
\end{center}
\end{figure}

In order to plot the parameter region (\ref{eq:para1})-(\ref{eq:para4}) in a two-plane, we normalize the parameters $a,j$ and $r_\infty$ as $A=a/\sqrt{m}, J=\sqrt{m}j$ and $R_{\infty}=r_\infty/\sqrt{m}$, respectively and furthermore, we fix the value of $R_\infty$. Then, in the cases of $R_\infty^2<2$, $R_\infty^2=2$ and $R_\infty^2>2$, the quadratic curve $R_\infty^4-2(1-4J(A+2J))R_\infty^2+2A^2=0$ in the condition (\ref{eq:para3}) becomes an ellipse, a line and a hyperbola,  respectively. The curve $R^2_\infty=1-4J(A+2J)$ in the condition (\ref{eq:para2}) has different shapes in the cases of $R_\infty^2<1$, $R_\infty^2=1$ and $R_\infty^2>1$. Hence we consider the cases of (i) $0<R^2_\infty<1$, (ii) $R_\infty^2=1$ (iii) $1<R^2_\infty<2$ (iv) $R_\infty^2=2$ and (v) $R_\infty^2>2$.
The shaded regions in FIG.\ref{fig:p1}-FIG.\ref{fig:p5} show the parameter region (\ref{eq:para1})-(\ref{eq:para4}) for a given $R_\infty$ in each case of (i)-(v), respectively.

\begin{figure}[!h]
  \begin{center}
   \includegraphics[width=0.8\linewidth]{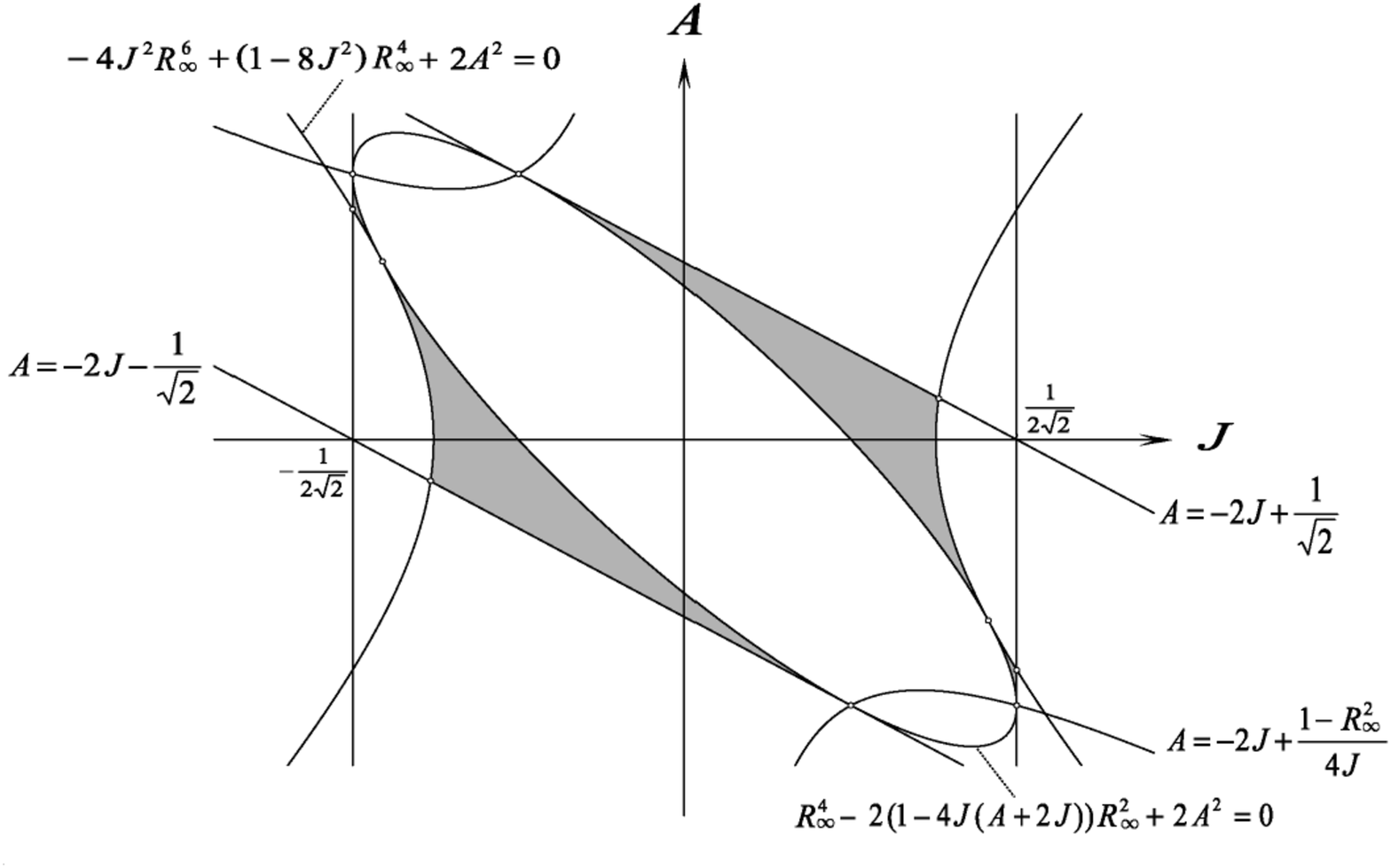}
  \caption{The parameter region in the $(J,A)$-plane in the case of 
 $1<R^2_\infty<2$.  \label{fig:p3}}
  \end{center}
\vspace{1cm}
\includegraphics[width=0.8\linewidth]{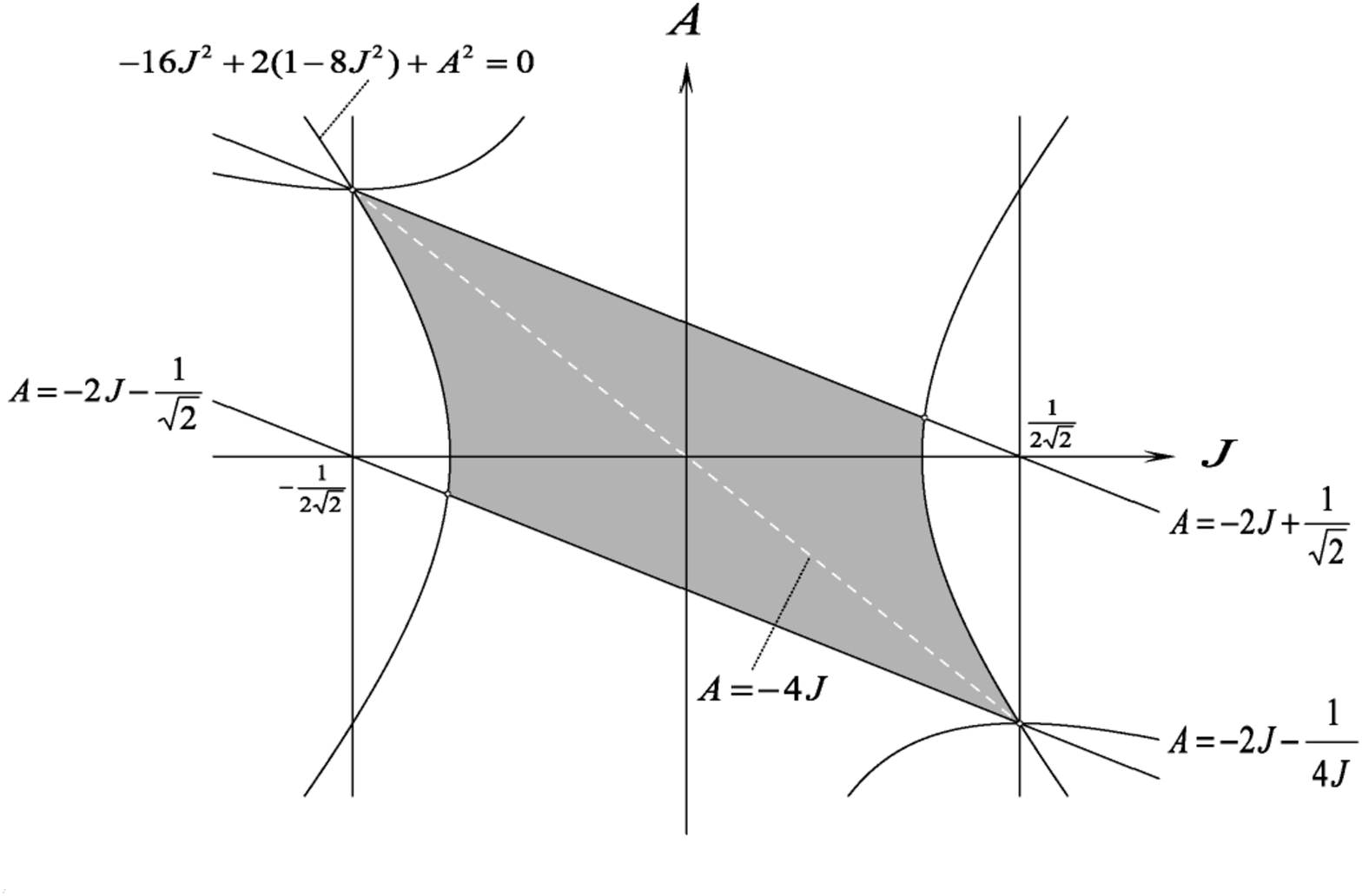}
\caption{The parameter region in the $(J,A)$-plane 
in the case of $R^2_\infty=2$.\label{fig:p4}}
\end{figure}

\subsection{Mass and angular momenta}
In the coordinate system $(t,r,\theta,\phi,\psi)$, the metric diverges at $r=r_\infty$ but we see that this is an apparent singularity and corresponds to the spatial infinity. To investigate the asymptotic structure of the solution, we introduce a new radial coordinate defined by
\begin{eqnarray}
\rho=\rho_0\frac{r^2}{r_\infty^2-r^2},
\end{eqnarray}
where the positive constant $\rho_0$ is given by
\begin{eqnarray}
\rho_0^2=\frac{r_\infty^2}{4} V(r_\infty).
\end{eqnarray}
Moreover, we introduce coordinates $(\bar t, \bar \psi)$ so that the metric is in a rest frame at the infinity
\begin{eqnarray}
\bar t=\frac{1}{C}t,\quad\bar \psi=\psi-\frac{D}{C}t,
\end{eqnarray}
where two constants $C$ and $D$ are chosen as
\begin{eqnarray}
C&=&\sqrt{\frac{2a^2m+r_\infty^4(1-4j^2(2m+r_\infty^2))}{2a^2m+8ajmr_\infty^2+r_\infty^2(-2m+16j^2m^2+r_\infty^2)}},\label{eq:A}\\
D&=&\frac{4(am+jr_\infty^4)}{\sqrt{\left(2a^2m+8ajmr_\infty^2+r_\infty^2(-2m+16j^2m^2+r_\infty^2)\right)\left(2a^2m+r_\infty^4(1-4j^2(2m+r_\infty^2)\right)}}\label{eq:B}.
\end{eqnarray}
Then, for $\rho\to \infty$, the metric behaves as
\begin{eqnarray}
ds^2&\simeq&-d\bar t^2+d\rho^2+\rho^2(\sigma_1^2+\sigma_2^2)+L^2\sigma_3^2,
\end{eqnarray}
where the angular coordinate $\psi$ in $\sigma_3$ is replaced by $\bar\psi$ and the size of an extra dimension, $L$, is given by
\begin{eqnarray}
L^2=\frac{2a^2m+r_\infty^4(1-4j^2(2m+r_\infty^2))}{4r_\infty^2}.
\end{eqnarray}
The Komar mass and the Komar angular momenta at the spatial infinity are given by
\begin{eqnarray}
M_{\rm K}&=&\frac{\pi}{\sqrt{\left(2a^2m+8ajmr_\infty^2+r_\infty^2(-2m+16j^2m^2+r_\infty^2)\right)(2a^2m+r_\infty^4(1-4j^2(2m+r_\infty^2))}}\nonumber\\
&\times& \biggl(-2a^2m^2+mr_\infty^4+32aj^3mr_\infty^4(m+r_\infty^2)+64j^4m^2(m+r_\infty^2)r_\infty^4+4ajm(2a^2m-r_\infty^4)\nonumber\\
&&+2j^2(-8m^2r_\infty^4-4mr_\infty^6+r_\infty^8+2a^2m(4m^2+4mr_\infty^2+3r_\infty^4)))\biggr),\\
J_{\psi}&=&-\frac{\pi}{2}\left[2a^2jm+2j^3r_\infty^6+m(-1+2j^2(4m+3r_\infty^2))a\right],\\
J_{\phi}&=&0,
\end{eqnarray}
respectively. 
Therefore the space-time has only the angular momentum in the direction of the extra dimension. In particular, it is noted that $J_\psi=\pi ma/2$ in the case of $j=0$ and $J_\psi=-\pi r_\infty^6 j^3$ in the case of $a=0$. Therefore, two parameters $(a,j)$ with the opposite signs means the rotations in the same direction, and  $(a,j)$ with the same signs means the rotations in the inverse direction.

\subsection{Near-Horizon Geometry and Regularity}
Here we investigate the near horizon geometry. As is mentioned previously, at the places such that $V(r)=0$, i.e., at $r=r_\pm$, the metric seems to be apparently singular but these correspond to Killing horizons.
To see this, we introduce new coordinates $(v,\psi')$ defined by
\begin{eqnarray}
&&t=v+\int \frac{2\sqrt{h(r)+r^2/4}k(r)}{rV(r)}dr,\\
&&\psi=\psi'+\int \frac{2k(r)g(r)}{rV(r)\sqrt{h(r)+r^2/4}}dr+\frac{g(r_\pm)}{h(r_\pm)+r_\pm^2/4}v.
\end{eqnarray}
In the neighborhood of $r=r_{\pm}$, the metric behaves as
\begin{eqnarray}
ds^2\simeq -\frac{r_{\pm}k(r_{\pm})}{2\sqrt{h(r_\pm)+r_\pm^2/4}}dvdr+\left[\frac{r_{\pm}^2}{4}k(r_{\pm})(\sigma_1^2+\sigma_2^2)+\left(h(r_\pm)+\frac{r_{\pm}^2}{4}\right)\sigma_3^2\right]+{\cal O}((r-r_{\pm})),
\end{eqnarray}
where the angular coordinate $\psi$ in $\sigma_3$ is replaced by $\psi'$. 
This means that the hypersurfaces $r=r_{\pm}$ are Killing horizons since the Killing vector field $\partial_v$ becomes null on $r=r_\pm$ and, furthermore, it is hypersurface-orthogonal, i.e., $V_\mu dx^\mu=g_{vr}dr$. We should also note that in the coordinate system $(v,\phi,\psi',r,\theta)$, each component of the metric form is analytic on and outside the black hole horizon. Hence the space-time has no curvature singularity on and outside the black hole horizon.

As will be explained later, the ergosurfaces are located at $r$ such that $g_{\bar t\bar t}=F(r^2)/r^4$ vanishes, where $F(r^2)$ is the cubic equation with respect to $r^2$ given by
\begin{eqnarray}
F(r^2)&=&-16j^2(am+jr^4_\infty)^2r^6\nonumber\\
&&-4(am+jr_\infty^4)[4a^2jm+am(-1+8j^2m)+jr_\infty^4(1-8j^2(m+r^2_\infty))]r^4\nonumber\\
&&-[2a^2m+r_\infty^4(1-4j^2(2m+r^2_\infty))]^2r^2+2mr_\infty^8[1-2j(a+2j(2m+r^2_\infty))]^2.
\end{eqnarray}

\section{Special cases}\label{sec:special}
In this section, we focus on a few simple cases of the solution.
\subsection{Rotating Gross-Perry-Sorkin (GPS) monopole}
First we consider the simplest case of $m=a=q=0$, where the solution has  no event horizon.
In terms of the coordinate $\rho$, the metric can be written as
\begin{eqnarray}
ds^2&=&-\left[dt+4j\rho_0^2\left(1+\frac{\rho_0}{\rho}\right)^{-1}\sigma_3 \right]^2\nonumber\\
    &&+\left(1+\frac{\rho_0}{\rho}\right)\left[d\rho^2+\rho^2(\sigma_1^2+\sigma_2^2)\right]+\rho_0^2\left(1+\frac{\rho_0}{\rho}\right)^{-1}\sigma_3^2,\label{eq:GPS}
\end{eqnarray}
where it should be noted that the requirement (\ref{eq:para4}) for the absence of CTCs imposes the parameters on the inequality $j^2<1/(16\rho_0^2)$.  
In the case of $j=0$, the (\ref{eq:GPS}) exactly coincides with the GPS monopole solution in Ref.\cite{GP}. It is noted that the point of $\rho=0$ is a fixed point of the Killing vector field $\partial_\psi$, and corresponds to a Kaluza-Klein monopole. This is often called a {\it nut}. In the presence of the parameter $j$, $\rho=0$ is also a fixed point of the Killing vector field $\partial_\psi$ and the metric is analytic at the point. Hence the presence of the parameter $j$ means that although the space-time has no black hole, it is rotating along the direction  $\partial_\psi$ of the extra dimension.
For $\rho\to\infty$, the metric behaves as
\begin{eqnarray}
ds^2&\simeq&-d\bar t^2+d\rho^2+\rho^2(\sigma_1^2+\sigma_2^2)+\rho_0^2\left[1-16j^2\rho_0^2\right]\sigma_3^2,
\end{eqnarray}
where it should be noted that CTCs at the infinity in the G\"odel universe vanish via the squashing transformation.

\begin{figure}[!h]
\begin{center}
\includegraphics[width=0.5\linewidth]{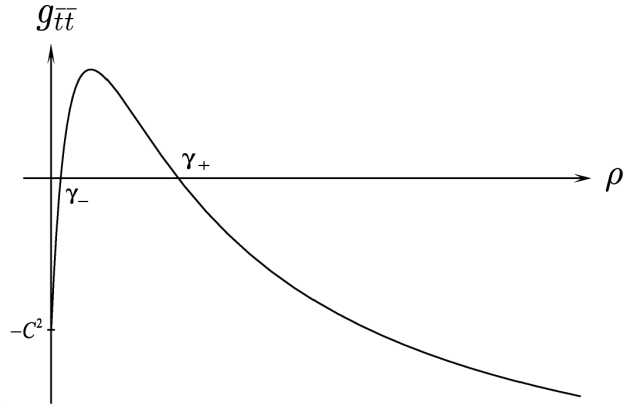}
\begin{minipage}{0.8\hsize}
\caption{The typical behavior of $g_{\bar t\bar t}$ in the case of 
$\sqrt{3}/8<|j|\rho_0<1/4$. There exists an ergoregion in the region 
such that $g_{\bar t \bar t}>0$. \label{fig:GPS_ergo}}
\end{minipage}
\end{center}
\end{figure}

Although this {\it rotating GPS monopole solution} has no event horizon, we can confirm the presence of an ergoregion.
The $\bar t\bar t$-component of the metric in the rest frame at the infinity take the form of
\begin{eqnarray}
g_{\bar t\bar t}=-\left[\left(C+\frac{1-C^2}{C}\left(1+\frac{\rho_0}{\rho}\right)^{-1}
\right)^2-\frac{1-C^2}{C^2}\left(1+\frac{\rho_0}{\rho}\right)^{-1}
 \right],
\end{eqnarray}
where the constant $C$ can be written in the form of $C=\sqrt{1-16j^2\rho_0^2}$ from Eq.(\ref{eq:A}).
As shown in FIG.\ref{fig:GPS_ergo}, in the case of $\sqrt{3}/8<|j|\rho_0<1/4$, $g_{\bar t \bar t}$ becomes negative in the region of $\gamma_-<\rho<\gamma_+$, where
\begin{eqnarray}
\gamma_\pm:=\rho_0\frac{1-3C^2\pm(1-C^2)\sqrt{1-4C^2}}{2C^2}.
\end{eqnarray}
 This means that there is an ergoregion, although there is no black hole horizon in the space-time and in the neighborhood of the nut, the ergoregion vanishes.

\subsection{Squashed Schwarzschild-G\"odel black hole}
\begin{figure}[htbp]
  \begin{center}
   \includegraphics[width=0.45\linewidth]{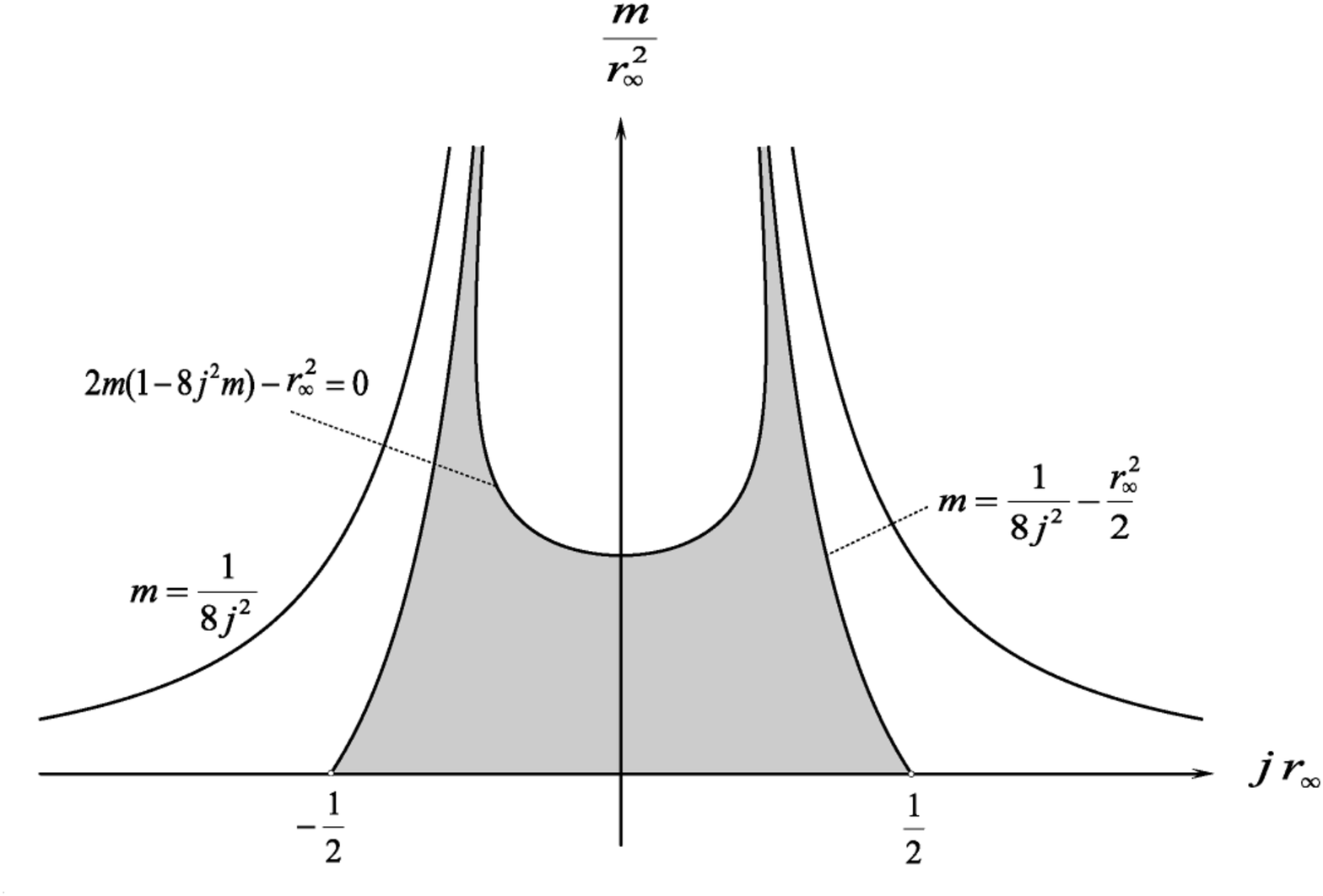}
 \begin{minipage}{0.8\hsize}
  \caption{Parameter region of the squashed Schwarzschild-G\"odel black hole solution in the $(j r_\infty,m/r_\infty^2)$-plane.}
  \label{fig:pararegion1}
 \end{minipage}
  \end{center}
 \vspace{0.7cm}
  \begin{center}
   \includegraphics[width=0.45\linewidth]{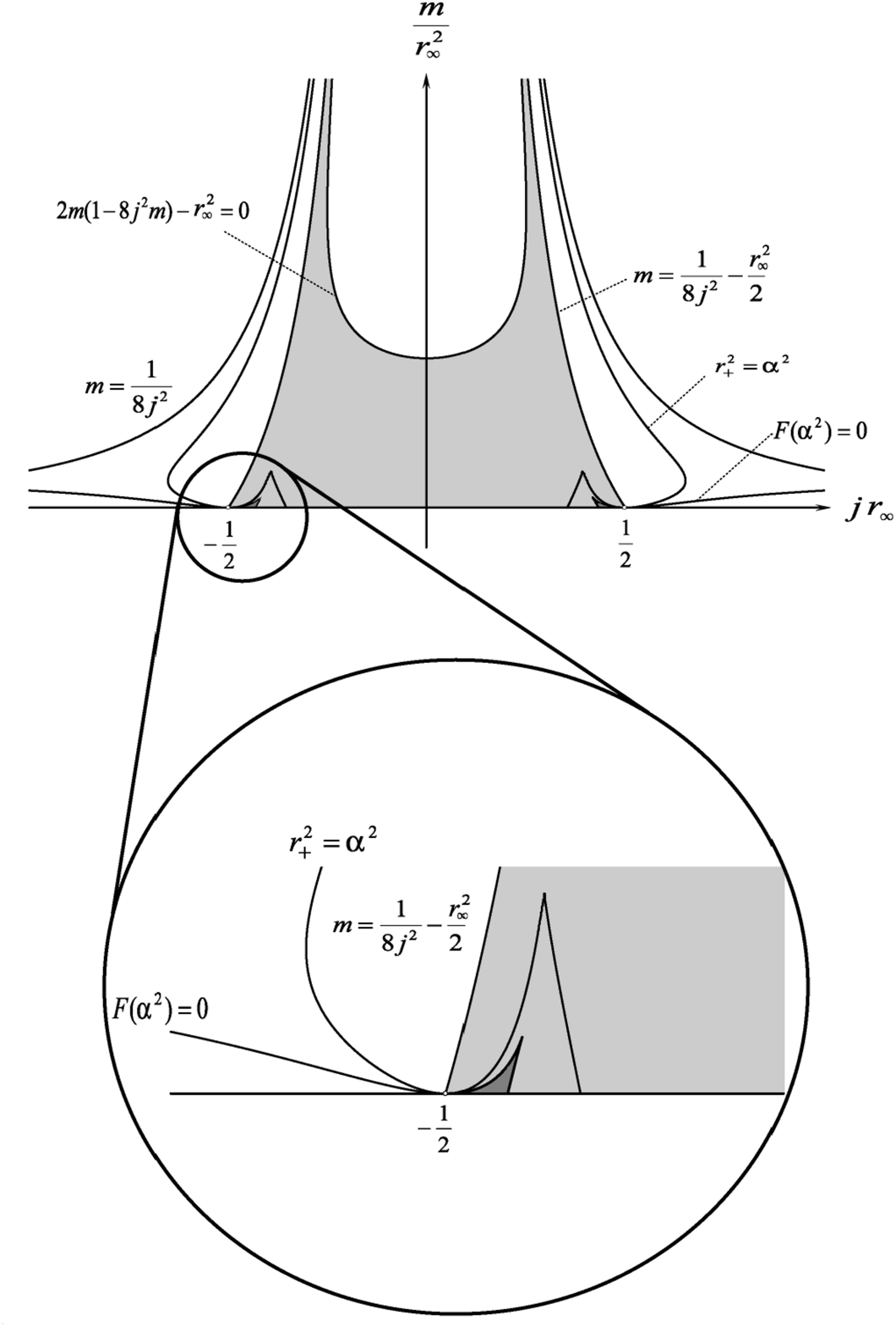}
  \end{center}
 \begin{minipage}{0.8\hsize}
  \caption{In the small dark parameter region in the above figure, there are two ergoregions. The below figure shows the close-up of the dark region in the above figure.}
  \label{fig:Sch_ergo}
 \end{minipage}
\end{figure}

Next, we consider the case of $a=q=0$. In this case, the solution is obtained via the squashing transformation for the Schwarzschild-G\"odel black hole solution in Ref.\cite{Gimon-Hashimoto}. The parameter region such that there is a black hole horizon and is no CTC outside the horizon becomes
\begin{eqnarray}
&&m>0,\label{eq:pp1}\\
&&1-8j^2m>0,\label{eq:pp2}\\
&&-2m+\frac{1}{4j^2}>r_\infty^2>2m(1-8j^2m)\label{eq:pp3}.
\end{eqnarray}
The horizon is located at $r_{\cal H}$ satisfying
\begin{eqnarray}
r_{\cal H}^2=2m(1-8j^2m).
\end{eqnarray}
The shaded region in FIG.\ref{fig:pararegion1} shows the region satisfying the inequalities (\ref{eq:pp1})-(\ref{eq:pp3}).

It should be noted that $g_{\bar t\bar t}(r=r_{\cal H})>0$ and $g_{\bar t\bar t}(r=r_\infty)<0$ always hold. Hence the ergosurfaces are always located at $r$ such that $g_{\bar t\bar t}=0$. In particular, when $F(\alpha^2)<0$ and $r_{\cal H}<\alpha$, remarkably there are two ergoregions $r_{\cal H}<r<r_1$ and $r_2<r<r_3$ outside the black hole horizon, where $\alpha^2$ and $\beta^2$ ($0<\alpha<\beta$) are the roots of the quadratic equation with respect to $r^2$,  $\partial_{r^2}F(r^2)=0$, and $r_i^2\ (i=1,2,3, \ r_1<r_2<r_3)$ are the roots of the cubic equation with respect to $r^2$, $F(r^2)=0$. The small dark region in FIG.\ref{fig:Sch_ergo} denotes a set of the solutions which admits two ergoregions outside the horizon.

\section{G\"odel parameter {\it versus} Kerr parameter}\label{sec:feature}
\subsection{Angular momentum}
\begin{figure}[!h]
\begin{center}
\includegraphics[width=0.8\linewidth]{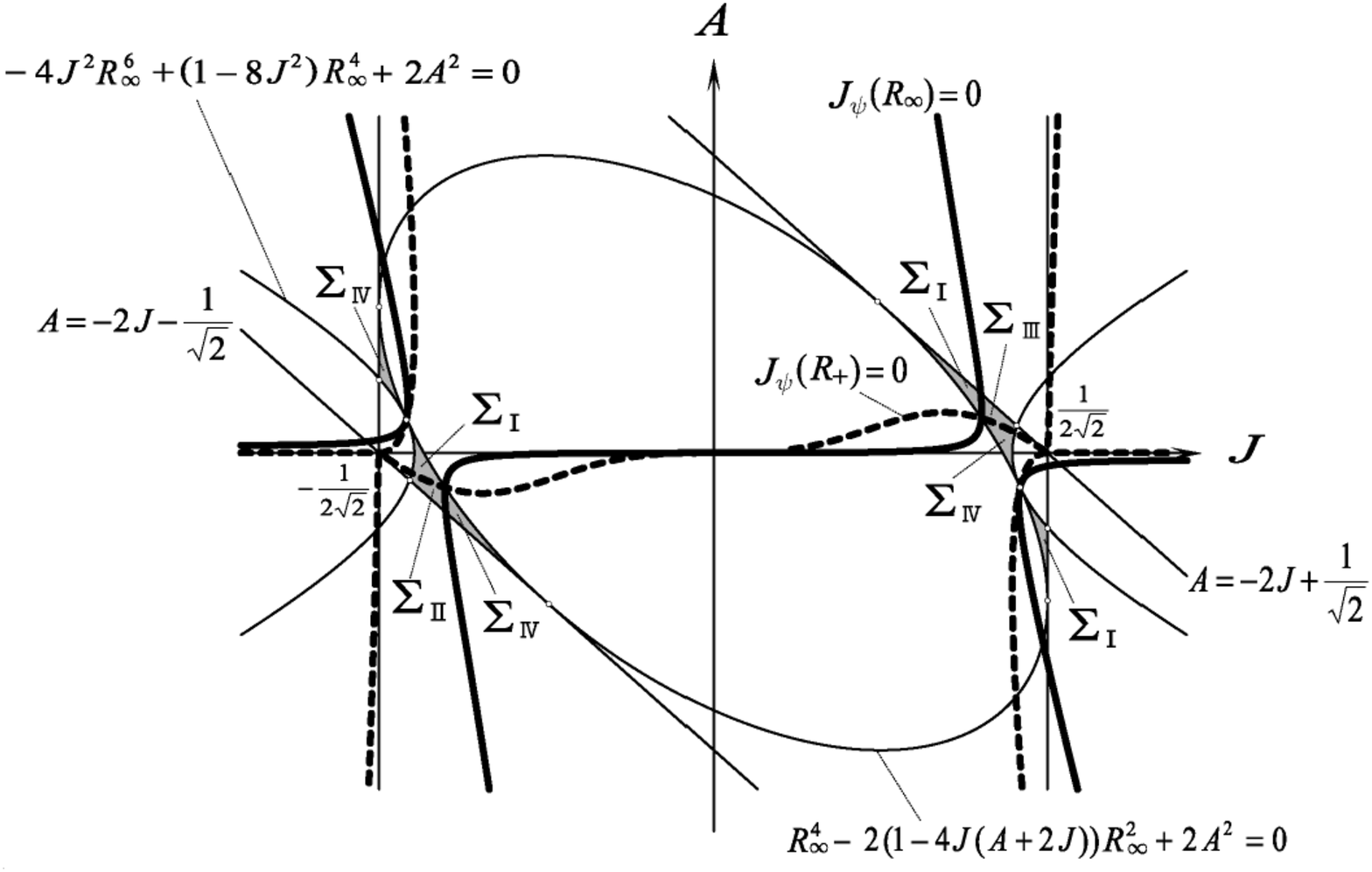}
\end{center}
\caption{The curves of $J_\psi(R_\infty)=0$ in the case of $0<R^2_\infty<1$. 
\label{fig:angularzero}}
\vspace{1cm}
\begin{center}
\includegraphics[width=0.8\linewidth]{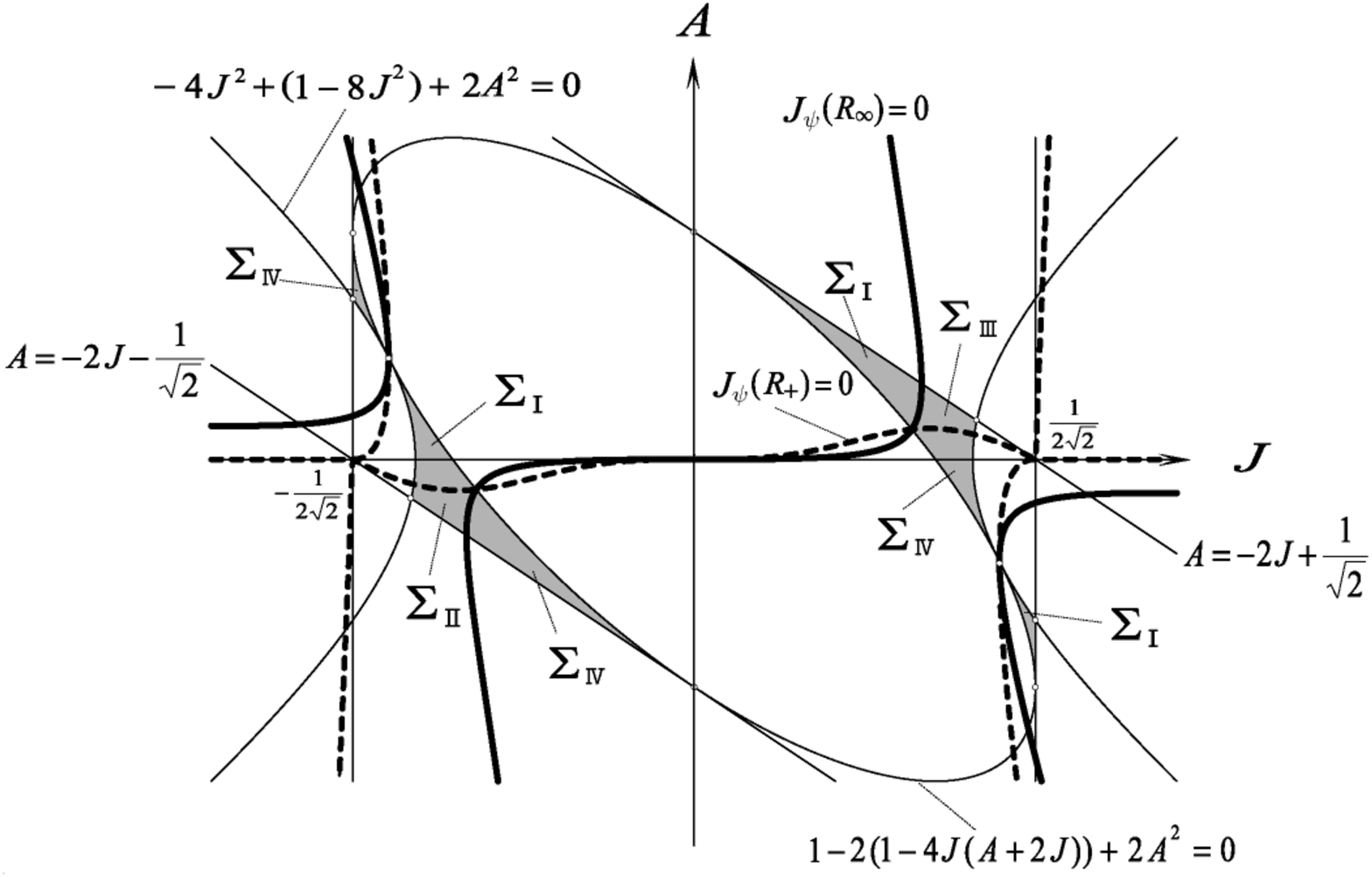}
\end{center}
\caption{The curves of $J_\psi(R_\infty)=0$ and $J_\psi(R_+)=0$ in the case of $R^2_\infty=1$. \label{fig:angularzero2}}
\end{figure}

\begin{figure}[!h]
\begin{center}
\includegraphics[width=0.8\linewidth]{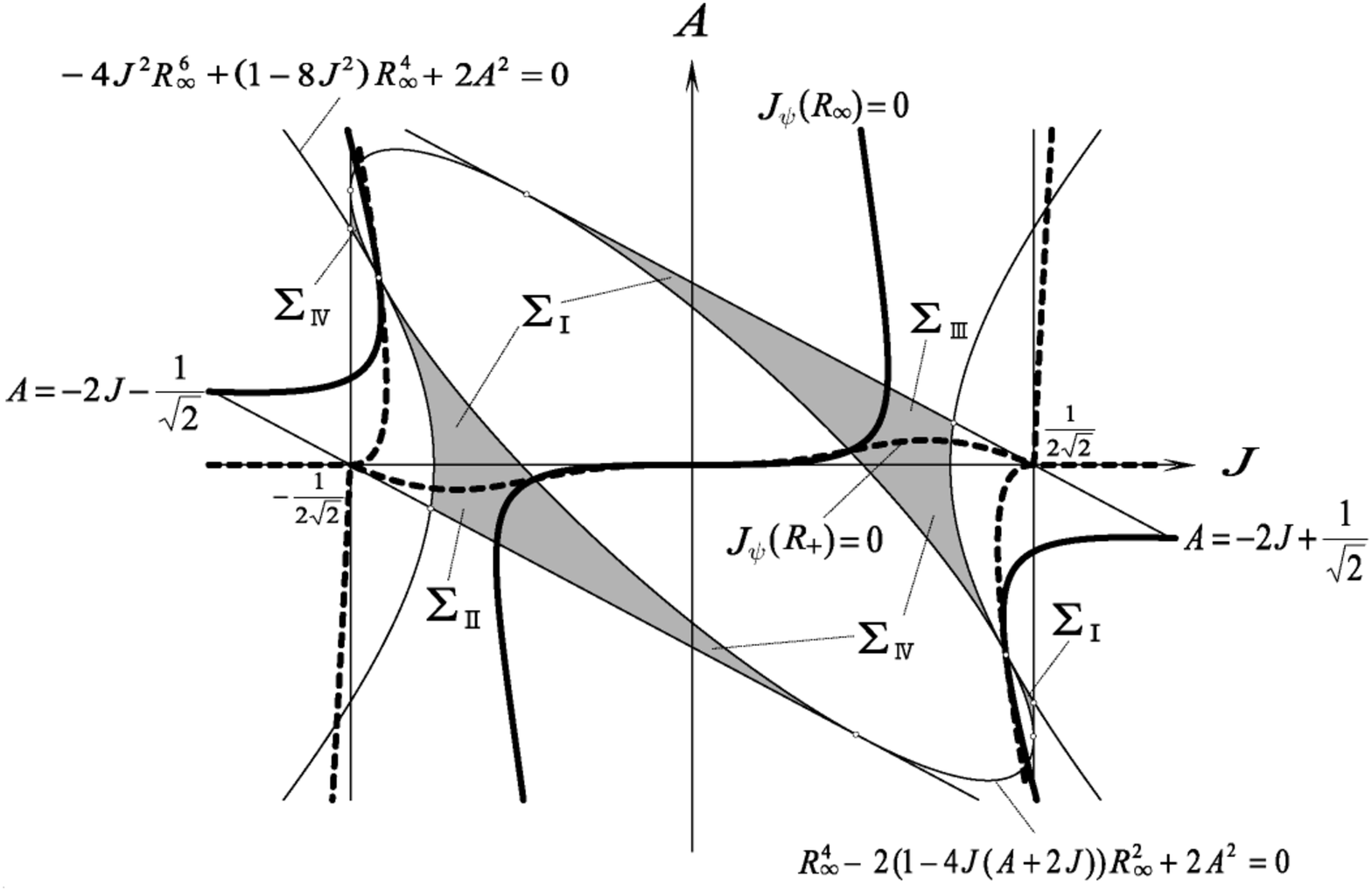}
\end{center}
\caption{The curves of $J_\psi(R_\infty)=0$ and $J_\psi(R_+)=0$ in the case of $1<R^2_\infty<2$. \label{fig:angularzero3}}
\vspace{1cm}
\begin{center}
\includegraphics[width=0.8\linewidth]{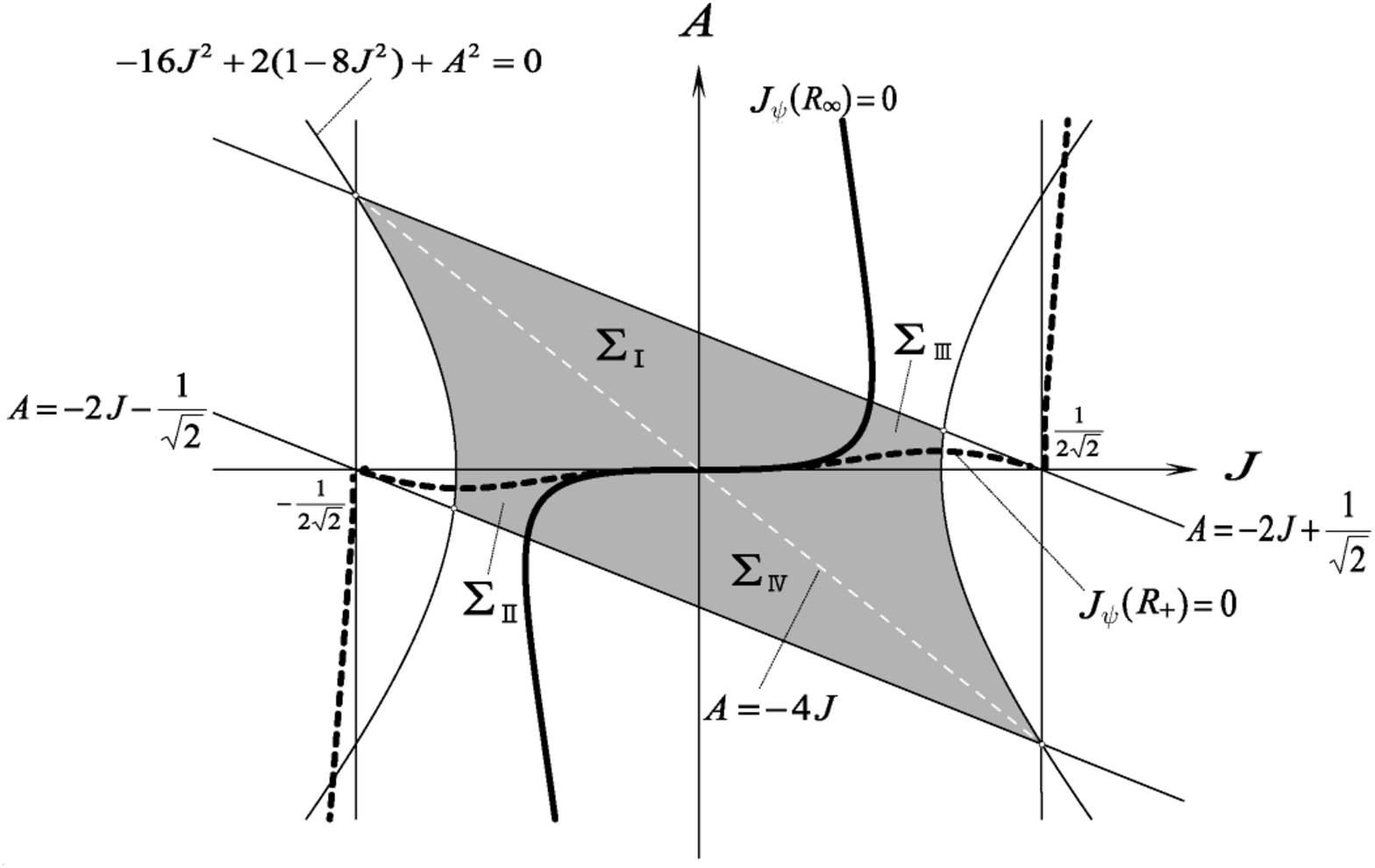}
\end{center}
\caption{The curves of $J_\psi(R_\infty)=0$ and $J_\psi(R_+)=0$ in the case of $R^2_\infty=2$. \label{fig:angularzero4}}
\end{figure}

\begin{figure}[!h]
\vspace{1cm}
\begin{center}
\includegraphics[width=0.8\linewidth]{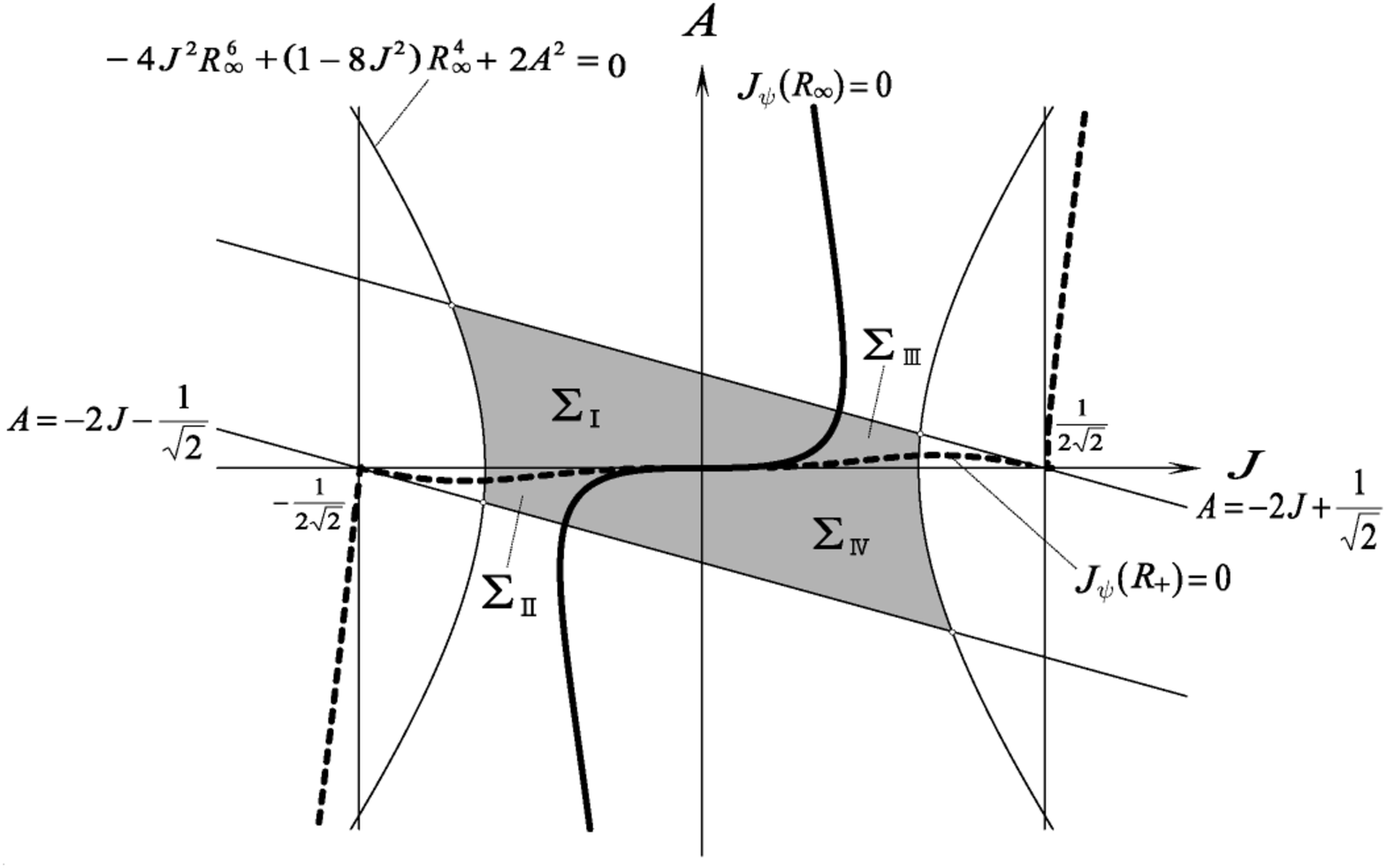}
\end{center}
\caption{The curves of $J_\psi(R_\infty)=0$ and $J_\psi(R_+)=0$ in the case of $R^2_\infty=3.0 $\label{fig:angularzero5}}
\end{figure}

The squashed Kerr-G\"odel black hole solution has two independent rotation parameters $a$ and $j$, where both $a$ and $j$ denote the rotation along the Killing vector $\partial_\psi$ (or $-\partial_\psi$) associated with a black hole and the (squashed) G\"odel universe, i.e., the rotating GPS monopole as the background, respectively. In this article, we call these parameters {\it Kerr parameter} and {\it G\"odel parameter}, respectively. Hence if the black hole is rotating in the inverse direction of the rotation of the (squashed) G\"odel universe, the effect of their rotations can cancel out the Komar angular momentum.  In fact, the total angular momentum with respect to the Killing vector field $\partial_\psi$, $J_{\psi}$, vanishes if the parameters satisfy
\begin{eqnarray}
2a^2jm+2j^3r_\infty^6+m(-1+2j^2(4m+3r_\infty^2))a=0.
\end{eqnarray}
As is shown in FIG.\ref{fig:angularzero}-\ref{fig:angularzero5}, the solid curve denotes a set of the solutions such that the total angular momentum $J_\psi$ vanishes for various values of $R_\infty$. In each case, as is expected, the curve of $J_\psi=0$ are located in the regions $A<0,J<0$ or $A>0,J>0$.
On the other hand, the Komar angular momentum $J_\psi(r)$ over the $r\ (r<r_\infty)$ constant surface is obtained as 
\begin{eqnarray}
J_{\psi}(r)=-\frac{\pi}{2}[2a^2jm+2j^3r^6+m(-1+2j^2(4m+3r^2))a].
\end{eqnarray}
The dotted curves in FIG.\ref{fig:angularzero}-\ref{fig:angularzero5} denote the curves of $J_\psi(R_+)=0$. The parameter region are decomposed into four regions $\Sigma_{\rm I}$, $\Sigma_{\rm II}$, $\Sigma_{\rm III}$  and $\Sigma_{\rm IV}$ by two curves $J_\psi(R_\infty)=0$ and $J_\psi(R_+)=0$ as follows
\begin{eqnarray}
\Sigma_{\rm I}&=&\{(J,A)\ |\ J_\psi(R_\infty)>0,\ J_\psi(R_+)>0\},\\
\Sigma_{\rm II}&=&\{(J,A)\ |\ J_\psi(R_\infty)>0,\ J_\psi(R_+)<0\},\\
\Sigma_{\rm III}&=&\{(J,A)\ |\ J_\psi(R_\infty)<0,\ J_\psi(R_+)>0\},\\
\Sigma_{\rm IV}&=&\{(J,A)\ |\ J_\psi(R_\infty)<0,\ J_\psi(R_+)<0\}.
\end{eqnarray}
In the solutions within $\Sigma_{\rm I}$ and $J<0$ or $\Sigma_{\rm IV}$ and $J>0$, $J_\psi(R_\infty)$ and $J_\psi(R_+)$ have the same sign since the black hole is rotating in the same direction as the squashed G\"odel universe (rotating GPS monopole). In the solutions within $\Sigma_{\rm I}$ and $J>0$ or $\Sigma_{\rm IV}$ and $J<0$, although two parameters $(J,A)$ has the same sign, $J_\psi(R_\infty)$ and $J_\psi(R_+)$ have the same sign because the effect of the black hole's rotation exceeds that of the G\"odel rotation. 

On the other hand, in the solutions within $\Sigma_{\rm II}$ or 
$\Sigma_{\rm III}$, the angular momenta $J_\psi(R_\infty)$ and $J_\psi(R_+)$ also have the opposite sign as well as $J$ and $A$. Therefore there exist the surface $R_0$ between the horizon $R=R_+$ and the infinity $R=R_\infty$ such that $J_\psi(R_0)$ vanishes. 
 In fact,  as is shown in FIG.\ref{fig:KG50p}, the angular momentum vanishes between the horizon and the infinity. It is expected that the effect of the G\"odel rotation exceeds that of the black hole's rotation at the large distance.

\subsection{Two counter rotating ergoregions}
\begin{figure}[!h]
  \begin{center}
   \includegraphics[width=0.7\linewidth]{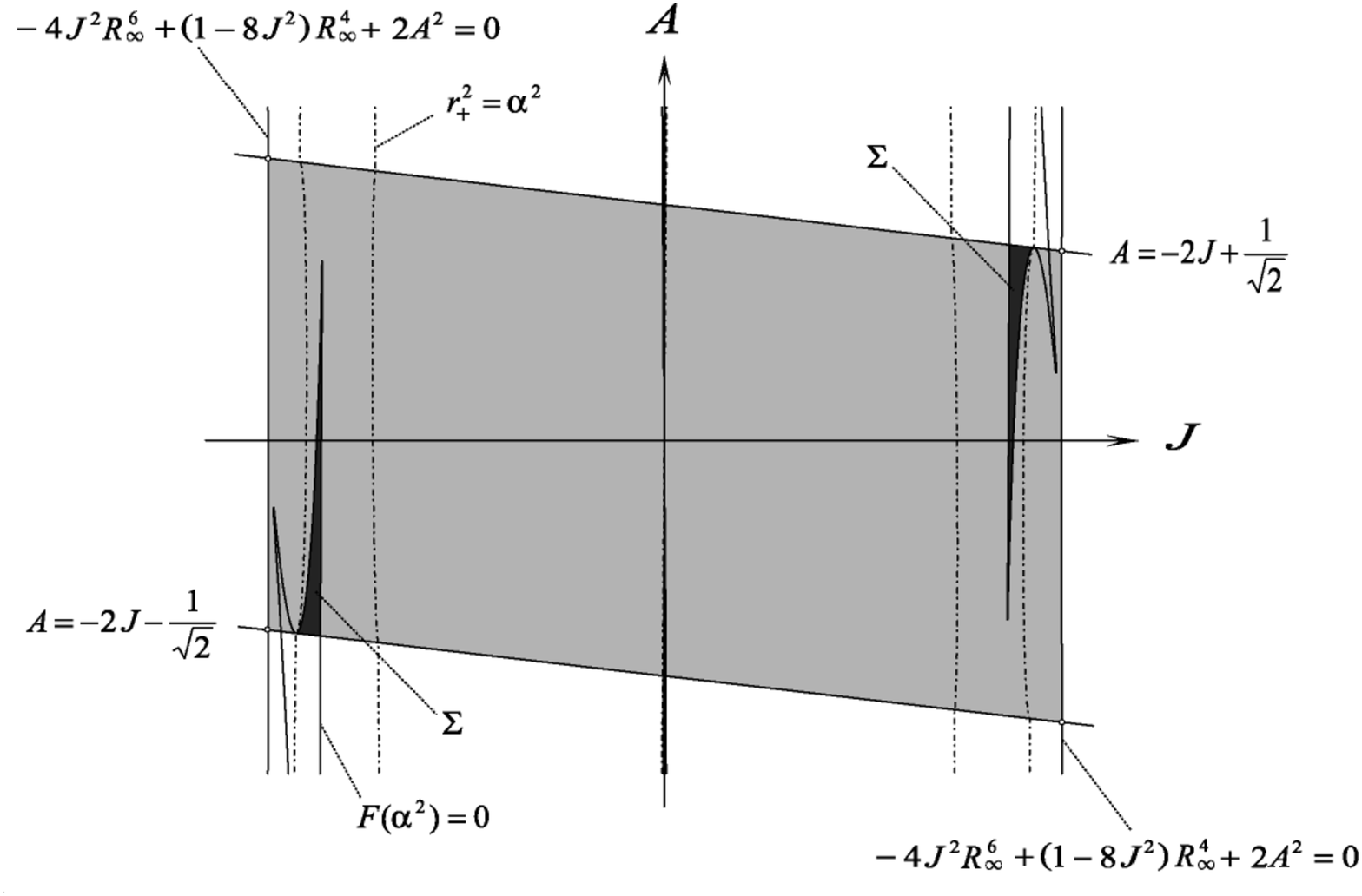}
  \end{center}
 \begin{minipage}{0.8\hsize}
  \caption{The small dark region $\Sigma$ shows the parameter region such that there are two ergo regions between the horizon and the infinity in the case of $R^2_\infty=50$}
  \label{fig:Komar_1}
 \end{minipage}
 \vspace{1cm}
  \begin{center}
   \includegraphics[width=0.7\linewidth]{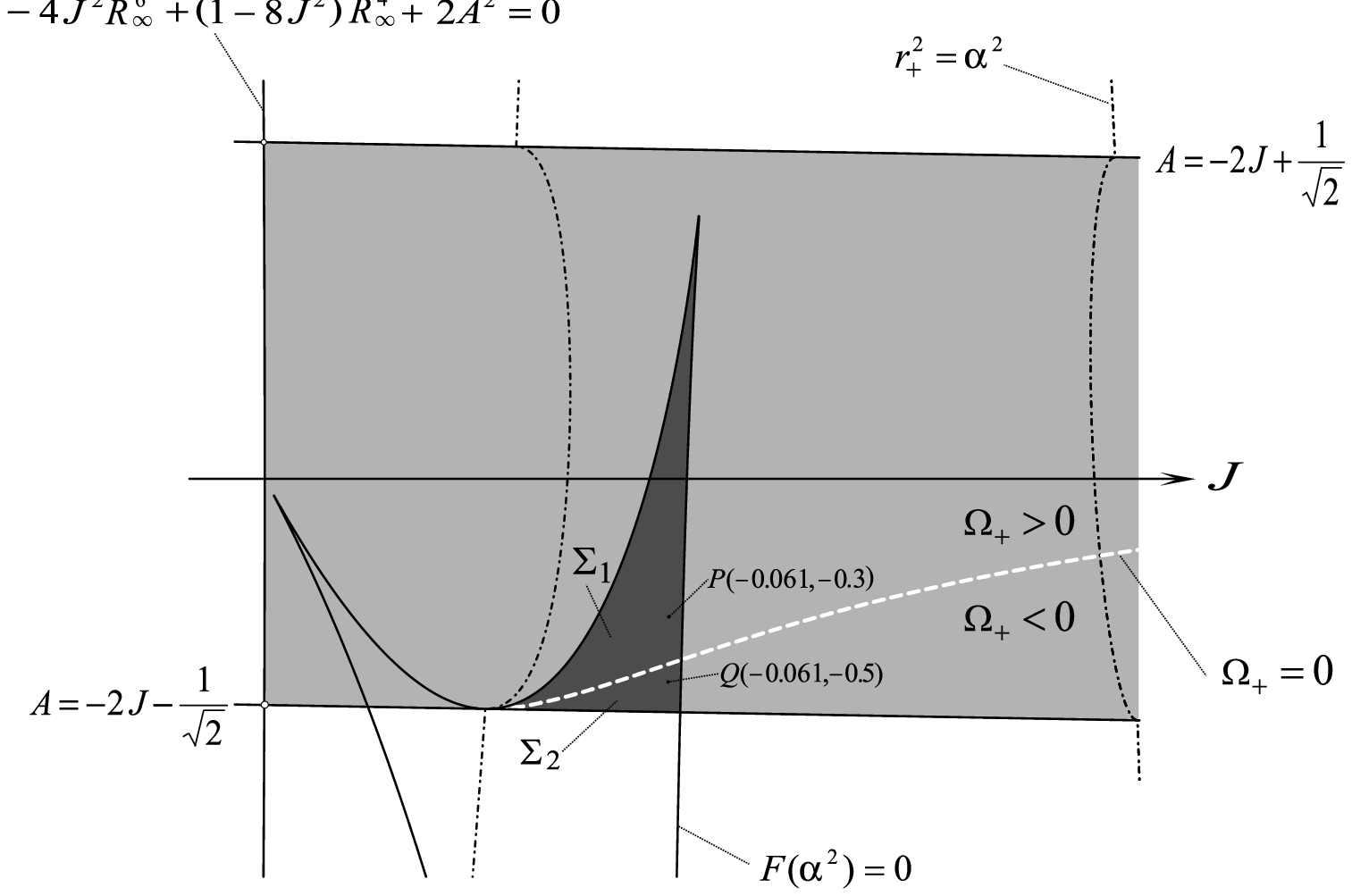}
  \end{center}
 \begin{minipage}{0.8\hsize}
  \caption{The close-up of the dark region in $J<0$ in FIG.\ref{fig:Komar_1}.}
  \label{fig:Komar_3}
 \end{minipage}
\end{figure}

Here we examine the ergoregions of the squashed Kerr-G\"odel black hole solution. The metric can be rewritten as follows 
\begin{eqnarray}
	ds^2&=&\left(h(r)+\frac{r^2}{4}\right)
		\left[\sigma_3+\left(-\frac{g(r)}{h(r)+\frac{r^2}{4}}C
	+D\right)d\bar t\right]^2\nonumber\\
&&	-\frac{\frac{r^2}{4}V(r)}{h(r)+\frac{r^2}{4}}d{\bar t}^2+\frac{k(r)^2}{V(r)}dr^2+\frac{r^2}{4}k(r)(\sigma_1^2+\sigma_2^2).
\end{eqnarray}
Therefore, at the outer horizon $r=r_+$ and the infinity $r=r_\infty$, the $\bar t\bar t$-component of the metric takes the non-negative form and the negative-definite form as follows
\begin{eqnarray}
&&g_{\bar t\bar t}(r=r_{+})=\left(-\frac{g(r_{+})}{h(r_{+})+r_{+}^2/4}C+D\right)^2\ge 0,\\
&&g_{\bar t\bar t}(r=r_\infty)=-\frac{r^2_\infty}{4}\frac{V(r_\infty)}{h(r_\infty)+r_\infty^2/4}C^2<0.
\end{eqnarray}
Hence, as is mentioned previously, the ergosurfaces are located at $r$ such that $g_{\bar t\bar t}=0$, i.e., the cubic equation with respect to $r^2$, $F(r^2)=0$. Like the squashed Schwarzschild-G\"odel black hole solution in the previous section, the space-time admits the presence of two ergoregions outside the black hole horizon. The spike like dark region $\Sigma$ in FIG.\ref{fig:Komar_1} correspond to the parameter region such that $F(r^2_+)>0$, $F(r_\infty^2)<0$, $F(\alpha^2)<0$ and $r_{+}<\alpha<\beta<r_\infty$ in the case of $R_\infty^2=50$.
It should be noted that in the case of $0<R_\infty^2\le2$, such a region does not appear within the parameter region (\ref{eq:para1})-(\ref{eq:para4}).

\begin{figure}[h]
  \begin{center}
   \includegraphics[width=0.8\linewidth]{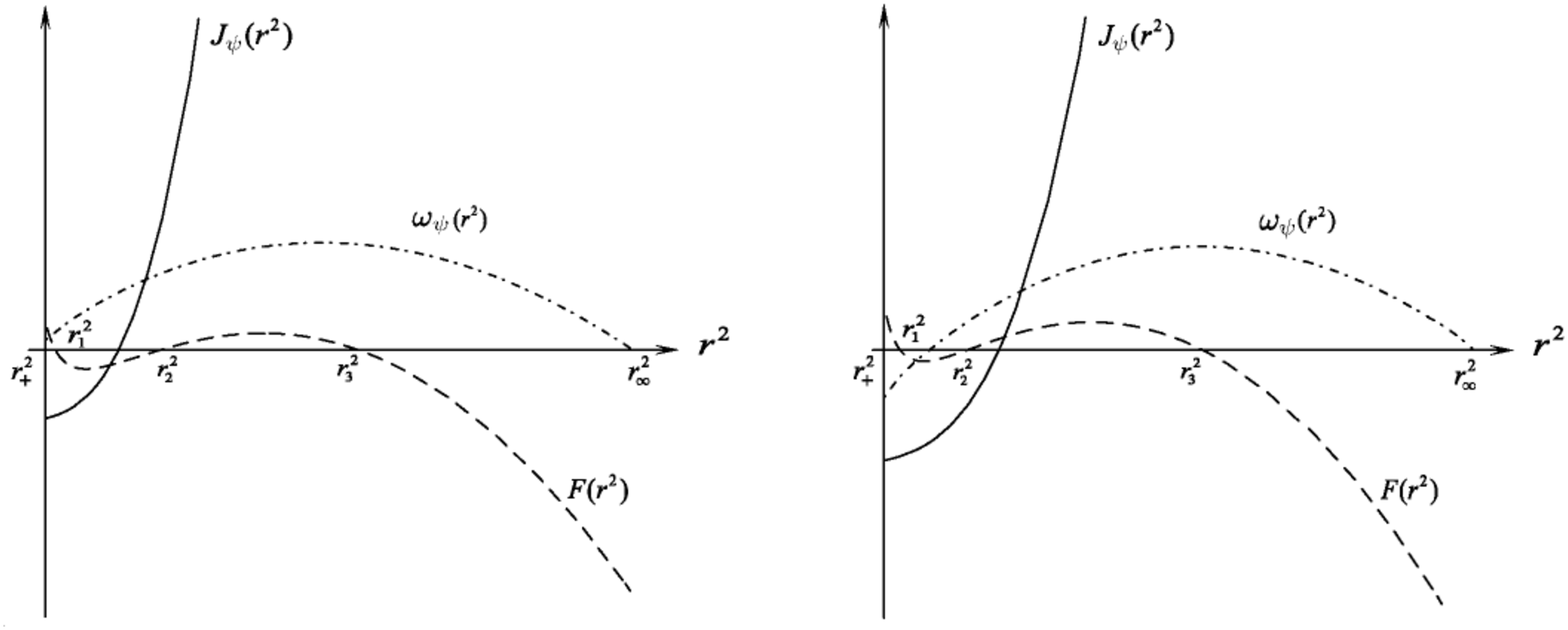}
  \end{center}
 \begin{minipage}{0.9\hsize}
  \caption{The left and right figures show the dependence on the radial coordinate $R$ of the angular velocity of a ZAMO $\omega_\psi(R)$ and the Komar angular momentum $J_\psi(R)$ in the case of $(A,J,R_\infty^2)=(-0.3,-0.061,50)$ and  $(A,J,R_\infty^2)=(-0.5,-0.061,50)$, respectively.}
  \label{fig:KG50p}
 \end{minipage}
\end{figure}

FIG.\ref{fig:Komar_3} shows the close-up of the dark region $\Sigma$ in FIG.\ref{fig:Komar_1}.
The dotted curve in FIG.\ref{fig:Komar_3} denotes $F(r_+^2)=0$, where the ergoregion near the outer horizon vanishes since in this case, the angular velocity $\Omega_+$ of the black hole horizon vanishes.
As is shown in FIG.\ref{fig:Komar_3}, the curve of $F(r_+^2)=0$ ($\Omega_+=0$) decomposes $\Sigma$ into two regions $\Sigma_1$ and $\Sigma_2$. In the region of $\Sigma_1$, the black hole horizon is rotating in the same direction as the G\"odel universe, while in the region of $\Sigma_2$, they are rotating in opposite directions. Here introduce virtual ZAMOs (zero angular momentum observer) located in the ergoregions.  The angular velocity of a ZAMO,  $\omega_\psi$, is given by $\omega_\psi:=-g_{\bar t \bar\psi}/g_{\bar \psi\bar \psi}$. At the horizon, note that $\omega_\psi|_{r=r_+}=\Omega_+$.
The dashed curves in FIG.\ref{fig:KG50p} denote the dependence on the radial coordinate $r$ of the angular velocity $\omega_\psi$ of ZAMOs in the case of $J=-0.061,A=-0.300,R_\infty^2=50$ in $\Sigma_1$ and $J=-0.061,A=-0.500,R_\infty^2=50$ in $\Sigma_2$, respectively. 
 In the region of $\Sigma_1$, a ZAMO in the ergoregion $r_+<r<r_1$ (ZAMO1) and a ZAMO in the ergoregion $r_2<r<r_3$ (ZAMO2) is rotating in the same direction $-\partial_\psi$, while  in the region of $\Sigma_2$, ZAMO1 and ZAMO2 are rotating in the direction of $\partial_\psi$ and $-\partial_\psi$, respectively.

\section{Summary and Discussions}\label{sec:summary}
Applying the squashing transformation to Kerr-G\"odel black hole solutions, we have constructed a new type of rotating Kaluza-Klein black hole solutions to the five-dimensional Einstein-Maxwell theory with a Chern-Simon term. The features of the solutions have been investigated. Though the G\"odel black hole solutions have closed timelike curves in the region away from the black hole, the new solutions generated via the squashing transformation have no closed timelike curve everywhere outside the black hole horizons. As the infinity, the space-time approaches a twisted $\rm S^1$ bundle over a four-dimensional Minkowski space-time. 

In the absence of a black hole, i.e., in the case of $m=a=0$, the solution describes the rotating GPS (Gross-Perry-Sorkin) monopole solution which is boosted in the direction of an extra dimension. Remarkably, in spite of the absence of a black hole, this space-time can have an ergoregion by the effect of the G\"odel rotation. In the squashed Schwarzschild-G\"odel black hole solution with the parameters $r_\infty,m$ and $j$, the space-time has one black hole horizon. It has one ergoregion around the black hole horizon or two disconnected ergoregions, i.e.,  an {\it inner ergoregion} located around the horizon and an {\it outer ergoregion} located far away from it. On the other hand, the squashed Kerr-G\"odel black hole solution with the parameters $r_\infty,m,a$ and $j$ describes a rotating black hole in the squashed G\"odel universe (the rotating GPS monopole background). The space-time has two horizons, i.e., an inner black hole horizon and an outer black hole horizon. The space-time has two independent rotations along the direction $\partial_\psi$ associated with the black hole and the squashed G\"odel universe. In the case of opposite rotations, the effect of the G\"odel's rotation and the black hole's rotation can cancel out the angular momentum at the infinity. In such a space-time, the ergoregions have richer and more complex structures.  Like the squashed Schwarzschild-G\"odel black hole solution, the space-time also admits the existence of two ergoregions, an inner ergoregion and an outer ergoregion. These two ergoregions can rotate in the opposite direction as well as in the same direction.

It should be noted that in addition to the mass and the angular momentum in the extra dimension, this space-time carries a electric charge $Q$ given by 
\begin{eqnarray}
Q:=\frac{1}{16\pi G_5}\int_{S^3}\left(\frac{2}{\sqrt{3}}*F-\frac{4}{3}A\wedge F\right)=2\sqrt{3}maj.
\end{eqnarray}
In particular, in the case of the rotating GPS monopole solution or the squashed Schwarzschild-G\"odel black hole solution, the electric charge vanishes. We can generalize our solution to a more general solution with  an extra parameter $q$ by applying the squashing transformation to a Kerr-Newman-G\"odel black hole 
solution in Ref.\cite{Wu}. The solution is given in Appendix\ref{app}. 
We leave the analysis of this solution for the future.

In the context of the study on dynamical features of BPS black holes,  black hole solutions 
on the Euclid base space~\cite{London,KS}, the Taub-NUT base space~\cite{IIKMMT} and on the Eguchi-Hanson base space~\cite{IMK} were also constructed in five-dimensional Einstein-Maxwell theory with a positive cosmological constant. It is expected that in the vanishing limit of the cosmological constant, the solutions describe the dynamics of BPS black holes. In particular, two-black hole solution on the Eguchi-Hanson space describes 
a non-trivial coalescence of black holes. 
In refs. \cite{IMK,YIKMT}, the authors compared the two-black hole solution 
on the Eguchi-Hanson space with the two-black holes solution 
on the Euclid space~\cite{London}, 
and discussed how the coalescence of five-dimensional black holes depends on 
the asymptotic structure of space-time. We also discussed the effect by rotation of black holes~\cite{MIKT}. In the future forthcoming article~\cite{MINT}, we will also study the dynamics in the BPS case of the squashed Kerr-Newman-G\"odel black hole solution by constructing such the solution in the five-dimensional Einstein-Maxwell theory with a Chern-Simon term and a positive cosmological constant.

\section*{Acknowledgments}
This work is supported by the Grant-in-Aid
for Scientific Research No.19540305. 

\appendix

\section{Squashed Kerr-Newman-G\"odel black holes}\label{app}
The metric and the gauge potential of the squashed Kerr-Newman-G\"odel black hole solution is given by
\begin{eqnarray}
ds^2=-f(r)dt^2-2g(r)\sigma_3dt+h(r)\sigma_3^2+\frac{k(r)^2dr^2}{V(r)}+\frac{r^2}{4}[k(r)(\sigma_1^2+\sigma_2^2)+\sigma_3^2],
\end{eqnarray}
and
\begin{eqnarray}
{\bm A}=\frac{\sqrt{3}}{2}\left[\frac{q}{r^2}dt+\left(jr^2+2jq-\frac{qa}{2r^2}\right)\sigma_3\right],
\end{eqnarray}
respectively, where the metric functions are 
\begin{eqnarray}
f(r)&=&1-\frac{2m}{r^2}+\frac{q^2}{r^4},\\
g(r)&=&jr^2+3jq+\frac{(2m-q)a}{2r^2}-\frac{q^2a}{2r^4},\\
h(r)&=&-j^2r^2(r^2+2m+6q)+3jqa+\frac{(m-q)a^2}{2r^2}-\frac{q^2a^2}{4r^4},\\
V(r)&=&1-\frac{2m}{r^2}+\frac{8j(m+q)[a+2j(m+2q)]}{r^2}\nonumber\\
    & &+\frac{2(m-q)a^2+q^2(1-16ja-8j^2(m+3q))}{r^4},\\
k(r)&=&\frac{V(r_\infty)r_\infty^4}{(r^2-r_\infty^2)^2}.
\end{eqnarray}
In the limit of $r_\infty \to \infty$, i.e., $k(r)\to 1$, the solution coincides with the Kerr-Newman-G\"odel black hole solution in Ref.\cite{Wu}.

\end{document}